\documentclass[fleqn,usenatbib]{mnras}
\usepackage{newtxtext,newtxmath}
\usepackage[T1]{fontenc}
\usepackage{ae,aecompl}

\usepackage[british]{babel}
\usepackage{graphicx}
\usepackage{amsmath}
\usepackage{xcolor}

\newcommand{\PC}{{\rm pc}}
\newcommand{\KPC}{{\rm kpc}}
\newcommand{\MSUN}{{\rm M}_\odot}

\newcommand{\MAG}{{\rm mag}}
\newcommand{\MICRON}{\mu{\rm m}}

\newcommand{\KMS}{{\rm km\,s}^{\rm -1}}

\newcommand{\MSUNSPC}{{\rm M}_\odot\,{\rm pc}^{-2}} 
 
\newcommand{\MSUNCKPC}{{\rm M}_\odot\,{\rm kpc}^{-3}} 
 
\newcommand{\V}{{\rm V}}

\newcommand{\M}{{\cal M}}
\newcommand{\SHI}{\Sigma_{HI}^{\scriptscriptstyle (thin)}}
\newcommand{\SmH}{\Sigma_{H_2}^{\scriptscriptstyle (CO)}}
\newcommand{\BI}[1]{{\it #1}}
\newcommand{\CHANGES}[1]{#1}	

\title[Baryonic masses of face-on spirals]{Estimating the baryonic masses of face-on spiral galaxies from stellar kinematics}
\author[F.V. Hessman]{Frederic V. Hessman \\
Institut f\"ur Astrophysik, University of G\"ottingen, Friedrich-Hund-Platz 1, 37077 G\"ottingen, Germany}

\date{Accepted 2017 April 12, Received 2017 April 11; in original form 2016 June 6.}

\pubyear{2017}

\begin{document}
\label{firstpage}
\pagerange{\pageref{firstpage}--\pageref{lastpage}}
\maketitle

\begin{abstract}
The kinematic dispersions of disc stars can be used to measure the dynamic contributions of baryons to the rotation curves of spiral galaxies and hence to trace the amount and distribution of the remaining dark matter. 
However, the simple single-component infinite disc model traditionally used to convert stellar dispersions to mass-densities is no longer adequate.
The dark matter halo has a significant effect upon the stellar dispersions for any non-maximal disc.
\CHANGES{The correction for cuspy dark matter halos is particularly large, suggesting that such models are not consistent with the observed stellar dispersions.}
When a more realistic model for the vertical gravity of the disc is used, the derived stellar surface densities are generally larger (smaller) for disc radii smaller (larger) than 2.3 times the radial scale-length. 
When the vertical gravity correction is applied to the radially resolved stellar mass-to-light ratios derived by the DiskMass consortium, the true values are not constant but decrease with radius, as expected from photometric colour gradients, and the true mass scale-lengths are about 80\% of the photometric scale-lengths.
The effects of a thin gaseous disc are larger than expected, especially when an allowance is made for optically thick or CO-dark gas. 
The presence of a thick-disc stellar component has severe consequences, particularly if its radial scale-length is smaller than that of the thin disc, as it appears to be in the Milky Way.
\end{abstract}



\section{Introdution}

The question of the relative amounts of baryonic and dark matter (DM) in galaxies is still one of the most pressing questions in extragalactic astrophysics, with broad impact on the question of the nature of DM and the role of baryons in the formation and evolution of galaxies 
\citep[e.g.]{2013MNRAS.429.3316B}. 
Traditionally, the amount of baryonic matter in the form of stars in the discs of spiral galaxies has been estimated photmetrically from assumptions about the mass-to-light ratios (M/L) \citep{2001ApJ...550..212B, 2009MNRAS.400.1181Z}, 
while the amount of gas has been estimated from HI and CO observations 
\citep{2013AJ....146...19L}. 
With robust baryonic surface densities, one can probe the DM distributions in the inner regions of galaxies, e.g. to answer questions about the cuspiness of DM halos 
\citep{2008IAUS..244...44W}. 
Unfortunately, the relative contributions of baryons and DM can be highly degenerate when fitting rotation curves.
The most obvious example of this phenomenon is the ``maximum-disc'' hypothesis: the inner discs of spirals (radii smaller than $\sim\!\!2.2\!$ times the radial scale-length $H_*$) are generally able to explain the inner rotation curves without the help of DM if one is allowed to scale the baryonic M/L as a fit parameter 
\citep{1986RSPTA.320..447V}. 
The latter is very difficult to estimate, however: the stellar light is due to a mixture of stars with different abundances, ages, and reddening 
\citep{2003MNRAS.344.1000B} 
and the results depend upon the stellar evolutionary model and spectral library \citep{2014PASA...31...36S, 2015MNRAS.449.2853R}. 
Within the Milky Way, this degeneracy can be broken by using stellar kinematics 
\citep{2013ApJ...779..115B},  
but only recently has it become possible to measure the kinematic dispersion of stellar populations within a reasonable fractional area of more distant spiral galaxies via 2-dimensional integral-field-unit (IFU) spectrographs 
\citep{2004PASP..116..565B, 2005PASP..117..620R, 2014ApJ...789...63A}. 
Since the dispersions of stars in thin discs are dominated by the local (presumedly mostly baryonic) surface densities, the study of the large-scale kinematics of stellar discs in relatively face-on spirals could help to reduce the systematic errors made in using photometry alone.

The {\it Disk Mass Survey} 
\citep{2010ApJ...716..198B, 2013A&A...557A.130M, 2013A&A...557A.131M} 
-- hereafter simply referred to as DiskMass or abbreviated as DMS --
was conceived as a means of determining the baryonic surface densities for a large number of nearby galaxies.
The goals were to confirm or disprove the maximum-disc hypothesis and to obtain robust baryonic surface densities by calibrating the stellar M/L independently of colour.
The basic procedure used by DiskMass was to measure the line-of-sight stellar dispersion from IFU spectroscopy, fit the dispersion with a model assuming constant ratios of the vertical, radial, and azimuthal components, estimate the vertical scale-height $h_*$ from the observed correlation with the easily measured photometric radial scale-length $H_p$ via the phenomenological relation
\begin{equation}
h_* \approx 200\,\PC \, \left(\frac{H_p}{\KPC}\right)^{0.63}
\label{eqn:h*}
\end{equation}
\citep{2010ApJ...716..234B} 
and to use a kinematic model to connect the observed (i.e. vertically-averaged and density-weighted) vertical dispersion $\overline{\sigma_z^2}(R)$ with the ``dynamic surface density''
\begin{equation}
\Sigma_{dyn}(R) \equiv \frac{ \overline{\sigma_z^2}(R) }{ \pi k G h_* } 
\label{eqn:dyn}
\end{equation}
\CHANGES{For the special case of an infinite disc with a single mass-component having an exponential vertical density distribution and scale-height $h_*$, this is equal to the true mass surface density if $k=3/2$}
\citep{1988A&A...192..117V}. 
Assuming that the dynamic surface density represents the total baryonic surface density, the stellar surface density $\Sigma_*$ can then be estimated by subtracting the contribution of the gaseous disc.
\begin{eqnarray}
\Sigma_{*}(R)
	& \! \approx \! &
	\Sigma_{dyn}(R) - \Sigma_{ISM}(R) 
	\\ \nonumber
	& \! \approx \! &
	\Sigma_{dyn}(R) - 1.4 \left(\Sigma_H(R) + \Sigma_{H_2}(R) \right)
\label{eqn:DiskMass}
\end{eqnarray}
The atomic gas contribution is derived from HI observations for the atomic gas; the molecular gas surface density $\Sigma_{mol}$ is assumed to follow the Spitzer $24\,\MICRON$ intensity assuming a connection between dust and CO 
\citep{2013ARA&A..51..207B, 2011ApJ...742...18W}. 
The factor of 1.4 corrects the pure hydrogen densities for helium and heavier elements.

The DiskMass approach to interpreting the stellar kinematics of face-on discs is, of course, not the only one.
For instance, one can use the much more complex method of 
\cite{2008MNRAS.390...71C}, 
in which the mass-distribution is modelled from the photometry as a sum of 3-dimensional Gauss functions from which the kinematic properties of stellar tracers can be derived from distribution functions, assuming a given M/L distribution.
While this is dynamically the ``better'' way of doing such an analysis, it too is fraught with difficulties due to the added complexity and assumptions.
\cite{2012ApJ...745...92A, 2014ApJ...789...63A} 
analysed similar IFU data for a smaller number of galaxies but derived the disc properties from the photometry assuming a constant M/L, after which a kinematic distribution function model was used mainly to constrain the DM halo properties; they were not able to obtain self-consistent potential solutions for the stellar dispersions and gas motions.
The DiskMass approach, while much simpler, has the advantage of attempting to derive disc properties with a minimum of theoretical assumptions and as close to the data as possible.

Although the DiskMass Survey has set new standards for the determination of the dynamic importance of baryons 
\citep{2015ApJ...801L..20C} 
using standard techniques and dynamical models,
there are several systematic problems associated with their procedures and assumptions beyond those discussed extensively in their series of papers.
For instance, their mass-to-light ratios, $\Upsilon_*$, are generally smaller than those derived using stellar population models 
\citep{2014AJ....148...77M}. 
DiskMass uses an empirical relation between the vertical and radial scale-lengths obtained from red and near-infrared images of edge-on galaxies, but 
\cite{2016MNRAS.456.1484A} 
show that this relation is likely dominated by the light of the older thick disc, resulting in an over-estimated vertical scale-height $h_*$ and a resulting under-estimated surface density $\Sigma_* \propto \sigma_z^2 / h_*$ and $\Upsilon_*$.
Eqn.\,\ref{eqn:DiskMass} was derived assuming that the gravity is due to a single mass-component with a given scale-height, i.e. either that both the ISM and the DM halo do not contribute or that they share the same vertical density distribution.
The former is obviously wrong -- the stellar surface density was derived thereafter by subtracting the ISM contribution from the dynamical surface density -- but in the latter case, the gaseous scale-heights are known to be significantly smaller and the density of DM is not thought to follow the stars at all.
Thus, the equation for $\Sigma_*$ used must be systematically incorrect at some level.
The equation also explicitly assumes that the local gravity is produced by the local surface density in an infinite disc: the non-local effects of gravity can substantially reduce or increase the local gravity over that of an infinite disc with constant surface density, depending upon the given radius, radial scale-length, and vertical scale-height 
\citep{1989MNRAS.239..571K}. 

The purpose of this paper is to identify additional systematic errors, formulate corrections to the procedure used by the Disk Mass Survey, and so to estimate the effects of the corrections on the derived stellar masses, M/L, and -- implicitly -- DM halo properties.


\section{Stellar kinematics and the dynamic surface density}

The equation we need to solve is the axially and vertically symmetric vertical Jeans equation connecting the dispersion tensor $\sigma_{ij}^{(k)}(R,z)$ and density $\rho^{(k)}(R,z)$ of some kinematic tracer population $(k)$\footnote{To avoid confusion, I will use superscript indices in parentheses to indicate kinematic tracers and subscript indices without parentheses to indicate mass-components.} with the local vertical gravitational acceleration $g_z(R,z)$
\begin{equation}
\frac{\partial}{\partial z} \left( \rho^{(k)} \sigma_{zz}^{(k)} \right) = \rho^{(k)} g_z(R,z) - \frac{1}{R} \frac{\partial}{\partial R} \left( R \rho^{(k)} \sigma_{Rz}^{(k)} \right)
\label{eqn:Jeans}
\end{equation}
and hence with the local total mass-density and potential.
The latter is determined by the (cylindrical) Poisson equation, 
\begin{equation}
-\frac{\partial g_z(R,z)}{\partial z} \equiv 4\pi G \rho(R,z) - \frac{1}{R}\frac{\partial}{\partial R} \left( R \, g_R(R,z) \right)
\label{eqn:Poisson}
\end{equation}
here expressed in terms of the accelerations\footnote{In the literature, these are often labeled $K_z$ and $K_R$ and are called (specific) ``forces''.} $g_i$ and the total mass-density $\rho$.

In the literature 
\citep{1984ApJ...276..156B, 1993A&A...275...16B}, 
the right-hand side of Eqn.\,\ref{eqn:Poisson} is often expressed in terms of 
\CHANGES{an effective ``baryonic''} density $\rho_b$ and an ``effective DM halo'' density using $|g_R| \! \approx \! V_c^2(R)/R$
\begin{equation}
\rho^{eff}_{halo} = \rho_{halo} + (4\pi G R)^{-1} \frac{\partial V_c^2(R)}{\partial R} 
\end{equation}
but the centripetal term actually applies to any potential with a radial gradient in $g_R$, e.g. also for realistic \CHANGES{baryonic} discs, 
\CHANGES{and so this separation is artificial and potentially dangerous.}

There are many possible approaches and simplifying assumptions one can make to solve these two equations.
All of the mass-components -- the stars, the ISM, and the DM halo -- principally have an effect, but the density of a roughly spherical halo may not be large enough to affect the vertical dispersion of a thin disc of stars even when it is otherwise dynamically dominant.
Thus, DiskMass used a simple, single-component infinite disc model to connect $\overline{\sigma_{zz}}(R) \equiv \overline{\sigma_z^2}(R)$ with the local mass surface density, a phenomenological model for the behaviour of the other diagonal components, $\overline{\sigma_R^2}(R)$ and $\overline{\sigma_\phi^2}(R)$, and neglected  other possible effects, some of which I will discuss in this section.

\subsection{The effects of the local gravity}

The classical model for $\overline{\sigma_z^2}$ used by DiskMass and others assumes that the local gravity responsible for maintaining a given level of vertical velocity dispersion is due to an infinite disc.
While this might be a reasonable assumption for a kinematically ``cold'' tracer, it can only be a crude approximation for the relatively geometrically thick, non-maximal and hence DM-dominated discs derived, e.g., by
\cite{2013A&A...557A.130M}. 
 
\cite{2016A&A...585A..17A} 
have criticised the simple approach for not including the effects of the DM halo.
If $h_*$ is the vertical scale-height, the stellar and DM contributions are roughly equal for
%
%
\begin{equation}
\rho_{DM,eq} \approx \frac{\Sigma_*}{2 h_*} \approx 3\,{\rm GeV/cm}^3 \left(\frac{\Sigma_*}{40\,\MSUN/\PC^3}\right) \left(\frac{h_*}{300\,\PC}\right)^{-1}
\label{eqn:DM}
\end{equation}
While this is a factor of $\sim \! 10$ larger than the generally accepted estimate of $\sim 0.3\,{\rm GeV/cm}^3$ for the local Milky Way 
\citep{2014JPhG...41f3101R}, 
the influence of the DM halo should be larger at larger radii and in galaxies with less stellar mass (e.g. dwarf galaxies).
In order to correct for this effect, Angus et al. assumed that $g_R(R,z)$ is roughly constant in height; when Eqn.\,\ref{eqn:Poisson} is vertically integrated, one obtains
\begin{equation}
|g_{z}(R,z)| \approx 2\pi G \Sigma(R,z) - |z| \frac{1}{R} \frac{\partial}{\partial R} \left( R \overline{g_R}(R) \right)
\label{eqn:AngusEtAl}
\end{equation}
where $\Sigma(R,z)$ is the integral of $\rho(R,z)$ between $\pm z$.
Adding mass-components which produce radial gradients in $g_R$ changes the dynamically active surface density -- and thereby the vertical dispersion -- and the sign of that change depends upon the sign of the gradient.
Dropping the $\sigma_{Rz}$ term in Eqn.\,\ref{eqn:Jeans} and assuming that the kinematic tracer $(k)$ has an exponential vertical distribution $\rho^{(k)} \propto e^{-|z|/h^{(i)}}$, Angus et al. used the vertically averaged dispersion
\begin{equation}
\overline{\sigma_{z}^2}\!(R)^{(k)} = \frac{1}{h^{(k)}} \int_{-\infty}^{+\infty} \int_{z}^{+\infty} e^{-|z'|/h^{(k)}} g_z (R,z') dz' dz
\label{eqn:complicated}
\end{equation}
to show that the inclusion of DM in $\Sigma$ and its corresponding effects on $g_z$ requires that there be less stellar mass than suggested by Eqn.\,\ref{eqn:DiskMass}, lowering the deduced DiskMass M/L even further.
However, their estimates were based on a cumbersome re-construction of the DiskMass data and mean M/L, so it is not obvious that the DM effects in the DiskMass analyses must be as dominant as they suggested.

The complicated forms of Eqn.\,\ref{eqn:AngusEtAl} and \ref{eqn:complicated} and of earlier formulations 
\citep{1984ApJ...276..156B, 1993A&A...275...16B} 
obfuscate some of the underlying effects.
An approach more similar to that of DiskMass but which at least corrects the implicit assumption that the kinematic dispersion is due to a single mass-component was made by
\cite{2013ApJ...772..108Z}. 
When the gravity is from an infinite stellar disc with an exponential vertical density profile with scale-height $h_*$, a geometrically thin gas disc, and a constant DM halo density $\rho_{DM}$, the solution to Jeans equation (minus the $\sigma_{Rz}$ term) is
\begin{equation}
\frac{\sigma_z^2(z)}{2\pi G h}
	\approx
	\Sigma_* \left(1\!-\!\frac{h_*}{h+h_*} e^{-|z|/h_*}\right) + \Sigma_{ISM}
	+ 2 \rho_{DM} (|z|\!+\!h) 
\label{eqn:z13}
\end{equation}
This equation is easily extended to include more mass components, since the surface densities enter through the vertical gravity only.
If one assumes that the kinematic tracer is identical with the stellar mass-component ($h \! \equiv \! h_*$) and integrates this expression over all heights $z$, weighted by the density, one obtains
\begin{equation}
\frac{\overline{\sigma_z^2}}{2\pi G h_*}
	~ \approx ~ 
	\frac{3}{4} \Sigma_* + \Sigma_g + 4\rho_{DM} h_* 
\label{eqn:iz13}
\end{equation}
(equivalent to integrating Eqn.\,\ref{eqn:complicated}).
When combined with Eqn.\,\ref{eqn:dyn}, an equation appears that resembles Eqs.\,\ref{eqn:dyn} and \ref{eqn:DiskMass} but where the effects of DM and the ISM have been included self-consistently:
\begin{equation}
\Sigma_{dyn} = \Sigma_* + \frac{4}{3} \Sigma_{ISM} + \frac{16}{3} \rho_{DM} h_*
\label{eqn:Sigma_dyn}
\end{equation}
The factor of $4/3$ before the gaseous surface density results from the assumption that the gaseous disc is much thinner than the stellar disc.
The effects of DM are also a factor $8/3$ stronger than we naively estimated in Eqn.\,\ref{eqn:DM}.

Eqn.\,\ref{eqn:Sigma_dyn} cannot be the entire story, since the gravity felt by the stars in a real galaxy is not that of an infinite disc.
\citep[Eqn.\,\ref{eqn:AngusEtAl}]{1989MNRAS.239..571K}.	
Fortunately, the non-infinite-disc effects in the simplified Jeans equation (Eqn.\,\ref{eqn:Jeans} minus the $\sigma_{Rz}$ term) enter only through $g_z(R,z)$ and are easily modelled as being due to sums of infinite-disc components with positive or negative effective surface densities and different vertical exponential scale-heights 
\citep{2015A&A...579A.123H}. 
This means that Zhang et al.'s simple infinite uniform disc model for the velocity dispersions is easily extendable for the case of a more realistic gravity field if one replaces the true local surface density and vertical scale-height with several effective infinite disc components.
For a two component model, the stellar term in Eqn.\,\ref{eqn:z13} can be replaced by
\begin{equation}
\Sigma_1 \left(1 - \frac{h_1}{h_*+h_1} e^{-z/h_1}\right)
+
\Sigma_2 \left(1 - \frac{h_2}{h_*+h_2} e^{-z/h_2}\right)
\end{equation}
After vertical averaging (weighted by the density), the stellar term in Eqn.\,\ref{eqn:Sigma_dyn} can then be written
\begin{equation}
\frac{4}{3} \left[ \left(1-\frac{h_1^2}{(h_*+h_1)^2} \right) \Sigma_1 +  \left(1-\frac{h_2^2}{(h_*+h_2)^2} \right) \Sigma_2 \right] \equiv \xi_* \, \Sigma_*(R)
\end{equation}
where $\xi_*$ scales the true to the effective stellar surface mass density and where the parameters $h_1$, $\Sigma_1$, $h_2$, and $\Sigma_2$ all depend upon $R$, $H_*$, and $h_*$.
However, since only ratios of scale-heights and surface densities are relevant, the correction is purely geometrical (independent of non-scaled disc properties), as expected.
Thus, we can define an effective surface density correction factor due to the vertically-averaged effects of the non-isotropic gravity from a doubly-exponential disc component $i$ with surface density $\Sigma_i(R)$,
\begin{equation}
\xi_i(R) ~ \equiv ~ \xi_i(R/H_i;h_i/H_i) ~ \equiv ~ \frac{\Sigma_{dyn,i}(R)}{\Sigma_i(R)}
\label{eqn:xi}
\end{equation}
as a measure of how the global density distribution of such a disc affects the vertically-averaged stellar velocity dispersion relative to that of an infinite disc with the local surface density.
Values of $\xi < 1$ imply that there is actually {\it more} stellar matter than one would naively estimate from the observed local dispersions and observed local surface densities, so $1/\xi$ is a measure of how much mass is actually present.

\begin{figure}
\includegraphics[width=\columnwidth, trim=0 1cm 0 1cm]{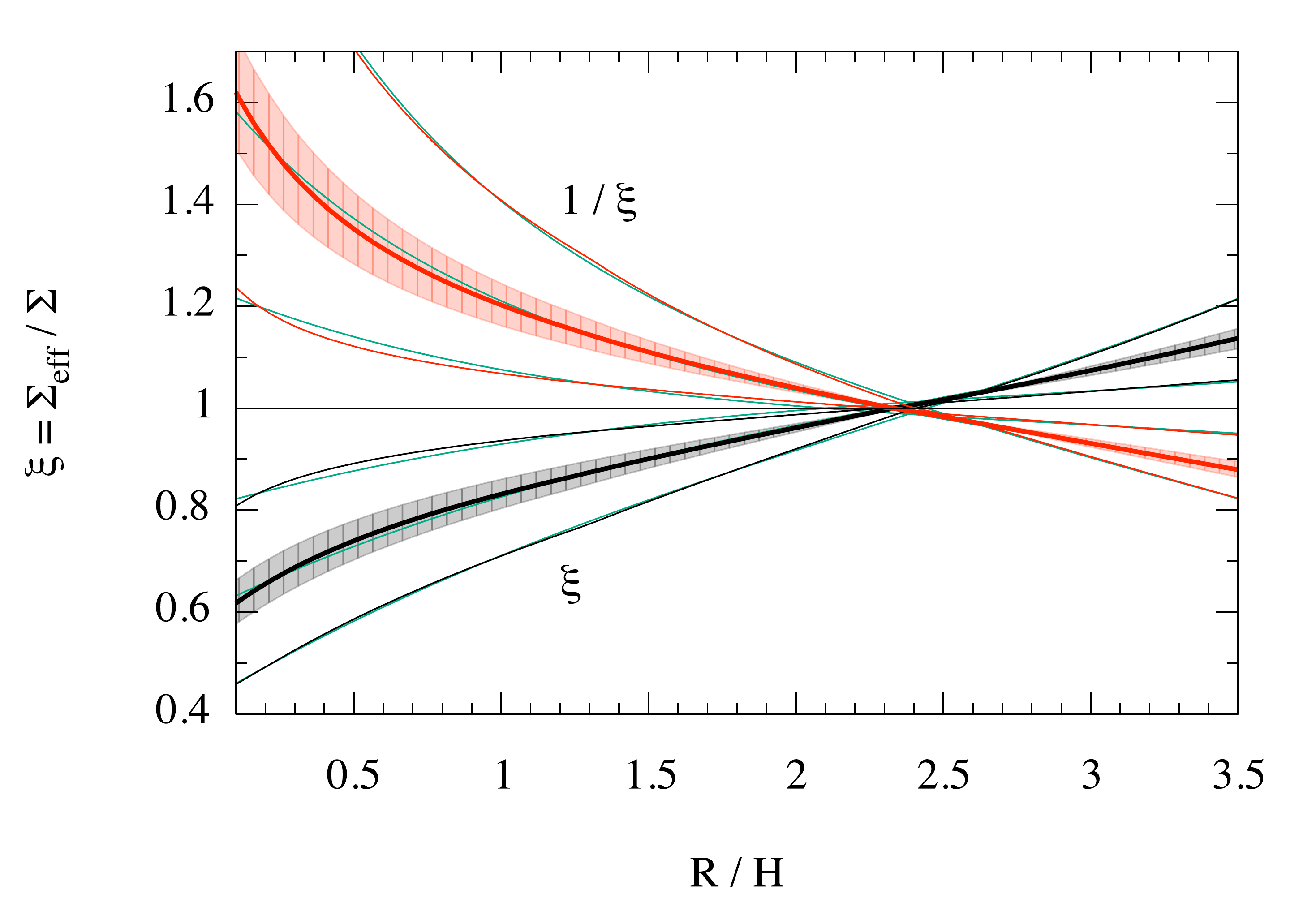}
\caption{The correction factor (Eqn.\,\ref{eqn:xi}, black lines) for the effective dynamic surface densities of a doubly-exponential disc with radial scale-length $H$ and constant vertical scale-height $h$ relative to those of a uniform infinite disc; the outer curves are for $h/H=0.04$ (shallowest) and 0.20 (steepest) whereas the bold line and shaded region corresponds to the average and range for the DiskMass galaxies (see text); the inverse value (red lines) is a measure of how much mass is underestimated using $\xi\!\equiv\!1$; the green lines are the polynomial fit given in Table\,1 and its inverse.}
\label{fig:xi}
\end{figure}

The result of calculating $\xi$ for various values of $R/H$ and $h/H$ is shown in Fig.\,\ref{fig:xi}: $\xi$ must be smaller in the inner disc, since the local surface density is untypically large; similarly, at some distance away from the center, the local exponential drop in surface density results in the local vertical gravity being determined more by the distant inner disc, resulting in a larger value of $\xi$.
The effect is negligible for $R/H \! \approx \! 2.3$ or for $h/H \! << \! 5\%$.
This effect, being purely geometric, is extremely robust, at least for a doubly-exponential disc and can be easily calculated using the simple polynomial approximation to $\xi(R)$ given in Table\,1 (shown as green lines in Fig.\,\ref{fig:xi}).
Note that this figure shows the vertically averaged effect for different $R/H$ whereas Fig.\,1 in 
\cite{2015A&A...579A.123H} 
shows the effect at a particular $R$; the extreme vertical gravities in the latter are for $R/H \! >> \! 2$ and $h/H \! > \! 2$, i.e. similarly extreme solutions in this Fig.\,\ref{fig:xi}.

The typical value of $\xi$ at the optical edge of a spiral galaxy can be derived assuming $R_{edge} \! \approx \! R_{25}$,
where $R_{25}$ is the radius at which the de-projected K-band surface brightness reaches 25$^{\MAG}$/square-arcsec:
\begin{equation}
\Sigma_{25} \equiv 0.87\,\MSUNSPC \Upsilon_K \equiv \Sigma_*(0) \, e^{-R_{25}/H_*}
\end{equation}
where $\Upsilon_K$ is the M/L of the stellar component in the K-band.
Using the average values of $H_K$ and $\log M_*$ for the DiskMass galaxies,
\begin{equation}
\frac{R_{25}}{H_*} \approx 5.6 + \ln\left[\left(\frac{M_*}{0.8\,10^{10}\,\MSUN}\right) \left(\frac{H_*}{4.2\,\KPC}\right)^{-2} \left(\frac{\Upsilon_K}{0.3}\right)^{-1} \right]
\end{equation}
$\xi$ thus reaches values of 1.4 or more near the typical optical edge of the disc.

Also shown in Fig.\,\ref{fig:xi} is the mean value and range of $\xi$ typical of the DiskMass galaxies with $0.09 < h_*/H_K < 0.13$ and $0.5 < R/H_K \! < \! 3$ 
\citep[their Table\,1]{2013A&A...557A.131M}. 
The mean correction to the effective dynamic surface density is small (-8\%) but the error produced by assuming $\xi \! \equiv \! 1$ results in a systematic gradient of $(1.1\!-\!0.74)/0.92 \! = \! 39\%$ is $\xi$ and $(0.93\!-\!1.4)/1.1 \! = \! -43\%$ in $1/\xi$ over the typical radial extent.
The fact that $\xi$ is systematically smaller in inner discs means that 
this geometric effect tends to make discs look less maximal than they really are.

The size of $\xi$ is determined by $h_*$ and $H_*$, the vertical and radial mass scale-lengths.
Note that the standard DiskMass procedure assumes that the photometric radial scale-length $H_p$ estimated from the surface brightness distribution in the K-band is essentially identical to that of the mass surface density.
If the true scale-length of $\Sigma_*$ is different
\begin{equation}
I(R) = I(0) e^{-R/H_p}, ~~~ \Sigma_*(R) = \Sigma_*(0) e^{-R/H_*}
\end{equation}
then $\Upsilon_*$ is not constant in radius and $H_*$ can be either larger or smaller than $H_p$
\citep{2014ApJ...797L..28S} 
-- an additional effect which will modify the derived amount of stellar mass present and which we will investigate in Section\,5.
Similarly, since $\Sigma_{dyn}$ is no longer directly proportional to $\Sigma_*$ (Eqn.\,\ref{eqn:Sigma_dyn}), there is no longer a direct causal reason why the radial scale-lengths of $\Sigma_*$ and $\overline{\sigma_z^2}$ should be identical; a difference would be reasonable and the observed similarity in the DiskMass dataset is no longer a guaranteed sign that the stellar component is totally dominant.

\begin{table}
\caption{Approximate polynomial fit to $\xi(R)$ (Eqn.\,\ref{eqn:xi})}
\begin{tabular}{l}
\hline
$\xi(u\!\equiv\!\frac{R}{H};v\!\equiv\!\frac{h}{H}) \approx a(v)\!+\!b(v)u\!+\!c(v)u^2\!+\!d(v) u^3$ \\
\hline
$a(v) = 0.9638    - 4.412  v + 12.67  v^2 - 20.27 v^3$ \\
$b(v) = 0.04554   + 3.482  v - 15.43  v^2 + 27.49 v^3$ \\
$c(v) = -0.01359  - 0.9580 v + 5.666  v^2 - 11.45 v^3$ \\
$d(v) = 0.0008014 + 0.1315 v - 0.7522 v^2 + 1.663 v^3$ \\
\hline
\end{tabular}
\end{table}

\subsection{The $\sigma_{Rz}$ cross-term}

The cross-term $\sigma_{Rz}$ produces additional effects in the Jeans equation that are visible in the spatially resolved kinematics of the Milky Way 
\citep{2015MNRAS.452..956B} 
and perhaps the Andromeda galaxy 
\citep{2016MNRAS.460.2720K}. 
The magnitude of the effect of the cross-term in the local Milky Way was estimated by 
\cite{2013ApJ...772..108Z} 
to be a factor of $\sim$0.01 smaller than the others.
In more distant galaxies, where only spectra integrated along the line-of-sight are available, it should be even more difficult to measure.

\cite{2016A&A...585A..17A} 
followed 
\cite{1989MNRAS.239..571K} 
and estimated the effect on the DiskMass results by assuming that the velocity ellipsoid points to the center of a galaxy, so that\footnote{Note that the definition of the DiskMass $\alpha$ parameter is exactly inverted relative to that used by 
\cite{1989MNRAS.239..571K}.}. 
\begin{equation}
\sigma_{Rz} \approx \frac{(\beta^{-2}-1)}{1+\frac{z^2}{\beta^2 R^2}} \frac{z}{R} \sigma_z^2
\end{equation}
where $\sigma_z \! \approx \! \beta \sigma_R$ (here, I have replaced Angus et al.'s $\alpha$ with the $\beta$ used by DiskMass).
They then calculated the effect by assuming a pure DM halo potential.
The result was a decrease in the vertical dispersions by 5-14\% at $R \! = \! H$, corresponding to an even slighter increase in the stellar M/L, and an even smaller effect at larger radii.

\cite{2015A&A...579A.123H} 
studied a simpler model for $\sigma_{Rz}$,
\begin{equation}
\label{eqn:sigmaRz}
\sigma_{Rz} \approx -\left(\beta^{-2}-1\right) \frac{z}{R} \sigma_z^2
\end{equation}
If $\sigma_z^2$ is roughly proportional to $e^{-R/L}$, as is seen in the DiskMass data 
\citep[Fig.12]{2013A&A...557A.130M}, 
Eqn.\,\ref{eqn:Jeans} can then be simplified to
\begin{equation}
\frac{\partial}{\partial z} \left( \rho^{(k)} \sigma_{zz}^{(k)} \right) 
\approx \rho^{(k)} \left[
g_z(R,z) + \sigma_{Rz}^{(k)} \left( \frac{1}{H^{(k)}}+\frac{1}{L^{(k)}} \right)
\right]
\label{eqn:ModelJeans}
\end{equation}
One sees that the $g_z$ and $\sigma_{Rz}$ terms have the same sign (e.g. both are negative for $z \! > \! 0$), so -- just as more gravity means more vertical dispersion for constant scale-height $h$ -- including the tilt-term tends to {\it increase} rather than {\it decrease} $\sigma_z^2$.
In turn, one generally needs {\it less} mass and {\it lower} M/L than when the term is ignored.
\cite{2016MNRAS.459.4191S} 
came to the same conclusion using a slightly simplified model where $\sigma_{Rz} \propto -z^n$ instead of $-z/R$; the difference between these conclusions and that of Angus et al. must then be due to the latter's particular treatment of the DM halo and modelled DiskMass data.

If the same simple DM and infinite disc gravity models used by 
\cite{2013ApJ...772..108Z} 
are kept, Eqn.\,\ref{eqn:ModelJeans} has an analytic solution 
\citep{2015A&A...579A.123H} 
that is easily density-weighted and integrated numerically over $z$ to obtain $\overline{\sigma_z^2}(R)$ for a particular set of galaxy parameters.
Note that a such a solution still depends upon modelling the non-uniform disc gravity using multiple effective infinite exponential mass components as discussed above.

Although these effects play a significant role in the vertically-resolved kinematics of the local Milky Way, one expects that the term is less important in the mid-plane regions with the highest exponential density-weighting and so must play a secondary role in spectra integrated along the line-of-sight.
We will see later, when full solutions for a typical DiskMass galaxy are constructed, that the final effects of the tilt-term are rather small in the outer discs, particularly when compared with other effects not included in the classical kinematic model, but can become significant in the inner discs.


\subsection{The kinematic effects of the ISM}

The ``razor-thin'' ISM disc plays a heightened role kinematically relative to the DiskMass assumptions via the additional factor of $4/3$ in Eqn.\,\ref{eqn:Sigma_dyn} for any given surface density.
In addition, DiskMass uses a simple measure for the neutral gas density based upon the assumption of optically thin HI and a molecular component derived indirectly from infrared and CO observations.
Detailed observations of the Milky Way and nearby galaxies have shown, however, that the cold neutral medium contains an optically thick component 
\citep{2005Sci...307.1292G, 2012ApJ...749...87B} 
that could have as much as 1.4 
\citep{2017MNRAS.464...32P} 
or even 2-3 times 
\citep{2014ApJ...796...59F} 
more mass than estimated using the standard methods.
There is also a CO-dark molecular component seen in $\gamma$-rays 
\citep{2012ApJ...755...22A} 
and other emission line tracers like C$^+$ 
\citep{2013A&A...554A.103P}. 
If one has to correct the total gaseous surface density upwards, the systematic effects on the kinematics are even larger.

\cite{2015A&A...579A.123H} 
and 
\cite{2015ApJ...814...13M} 
discuss the possible corrections and their kinematic effects within the context of the local Milky Way.
McKee et al.\ adopted a 30\% correction for the amount of neutral hydrogen over that assuming optically thin emission due to 
\cite{2012ApJ...749...87B}, 
but as a beam-smearing correction, this is a lower limit.
When discussing the important conversion of CO-emissivity to H$_2$ density calculated by 
\cite{2013ARA&A..51..207B}, 
McKee et al.\ say ``... that this empirical conversion factor includes the H$_2$ that lies in surface layers that have little CO but mostly C$^+$, the so-called {\it dark} or {\it hidden} H$_2$''.
However, they are actually referring to still crude (if admittedly non-trivial) corrections made using models for the dust emissivity, the structure of photodissociation regions, the assumed UV radiation field strengths, and for the uniformity of cosmic-ray fluxes, all of which cannot really be considered definitively  ``empirical'' given the complex nature of the ISM.
Thus, reasonable estimations of these effects result in an increase in the neutral and perhaps in the molecular surface densities over those used by DiskMass by factors of $\sim\!1.3$, another effect which naively make the discs more maximal by adding more baryons.

The effects of under-estimated ISM densities can be included by using the correction factors $\zeta_{HI}$ and $\zeta_{H_2}$; they undoubtedly have their own radial, metallicity, and other dependences 
\citep{2016MNRAS.462.2804N},  
but one can initially assume they are constant.
A new definition for the dynamic surface density that includes all of the additional effects DiskMass should have included is then
\begin{eqnarray}
\Sigma_{dyn}(R) \!
	& \!\approx\! &
	\xi_*(R) \Sigma_*(R) + \frac{16}{3}\rho_{DM}(R) h_*
	\\ \nonumber
	& & ~~~~~
	+ \frac{4}{3} 1.4 \left( \zeta_{HI} {\SHI}(R) + \zeta_{H_2} {\SmH}(R) \right)	
\label{eqn:eqn}
\end{eqnarray}
Here, I have implicitly assumed that the correction for the anisotropic gravity of the gaseous discs, $\xi_{HI}$ and $\xi_{H_2}$, can be ignored; the vertical scale-heights are smaller and the radial scale-lengths of the total ISM are typically twice those of the stars 
\citep{2012ApJ...756..183B}, 
resulting in a much smaller effect.


\section{Stellar masses and M/L ratios}

The DiskMass Survey assumed that the stellar density can be derived by simply subtracting their estimated gaseous surface densities even though this procedure is formally inconsistent with their definition of the dynamic surface density for ISM and DM contributions not distributed as the stars.
Using the modified Zhang et al. model and assuming a single stellar component, one obtains instead
\begin{eqnarray}
\Sigma_*(R) \! \! \!
	& \! \approx \! & \! \!
	\left[ \frac{1}{\xi_*(R)} \right] \Sigma_{dyn}(R)
	- \left[ \frac{16 \rho_{DM}(R) h_*}{3 \xi_*(R)} \right]
	\\ \nonumber
	& & \! \!
	- 1.4 \left(
		\left[ \frac{4 \zeta_{HI}(R)}{3 \xi_*(R)} \right] {\SHI}(R)
	+	\left[ \frac{4 \zeta_{H_2}(R)}{3 \xi_*(R)} \right] {\SmH}(R) 
		\right)
\label{eqn:Sigma*}
\end{eqnarray}
%
where the quantities in the square brackets have been used to emphasise the differences from Eqn.\,\ref{eqn:DiskMass} 
and are assumed to be equal to unity (baryons) or zero (DM) in the DiskMass analyses.

Given the typical behaviour of $\xi_*(R)$ (Fig.\,\ref{fig:xi}) 
and assuming values of $\zeta_{HI} \! \approx \! \zeta_{H_2} \! \approx \! 1.3$. one can qualitatively estimate the systematic errors in the DiskMass stellar densities and hence the M/L.
In the inner discs, where $1/\xi_* \approx 140$\%, the dynamic contribution to $\Sigma_{dyn,*}$ was significantly under-estimated and there should be more stellar mass and a higher M/L.
However, the gas there is mostly molecular and $4 \zeta_{H_2}/3 \xi_* \approx 290$\% so the gaseous correction was off by a large factor, potentially decreasing the stellar masses and stellar M/L dramatically, depending upon how gas-rich the inner disc is.
In the outer discs where $1/\xi_* \approx 90$\%, the dynamic contribution was slightly over-estimated, suggesting less stellar mass and lower M/L.
The outer ISM correction was also too small by an only slightly smaller factor of $4 \zeta_{HI}/3 \xi_* \approx 160$\%.
Thus, \BI{the derived DiskMass stellar M/L for the inner discs tend to be systematically smaller for gas-poor inner galaxies.}
The effects of the gas depend upon the relative surface densities, but then tendency is for the and systematically larger for gas-rich inner galaxies.
Thus, \BI{the DiskMass M/L tend to be too large for both gas-poor and gas-rich outer galaxies.}

These estimate were made, however, assuming that the ISM surface densities and their correction factors -- assumed to be constant -- are correct.
\cite{2013A&A...554A.103P} 
showed that the amount of ``CO-dark'' molecular gas in the Milky Way is lower in the inner disc (their Fig.\,18), somewhat alleviating the problem with the large molecular gas correction, perhaps to a still-large factor of $4/3 \xi_* \approx 130-190$\%.
However, they also showed that the cold neutral medium responsible for $\zeta_{H_2}$ represents a substantially larger (smaller) fraction of the neutral ISM of the Milky Way for Galactic radii smaller (larger) than the Sun's, where the fraction is roughly 40\% (their Fig.\,13).
If the same is true in external galaxies, this implies that the atomic correction is much larger in the inner discs, exacerbating the problem there again.

\begin{figure}
\includegraphics[width=\columnwidth, trim=0 1cm 0 1cm]{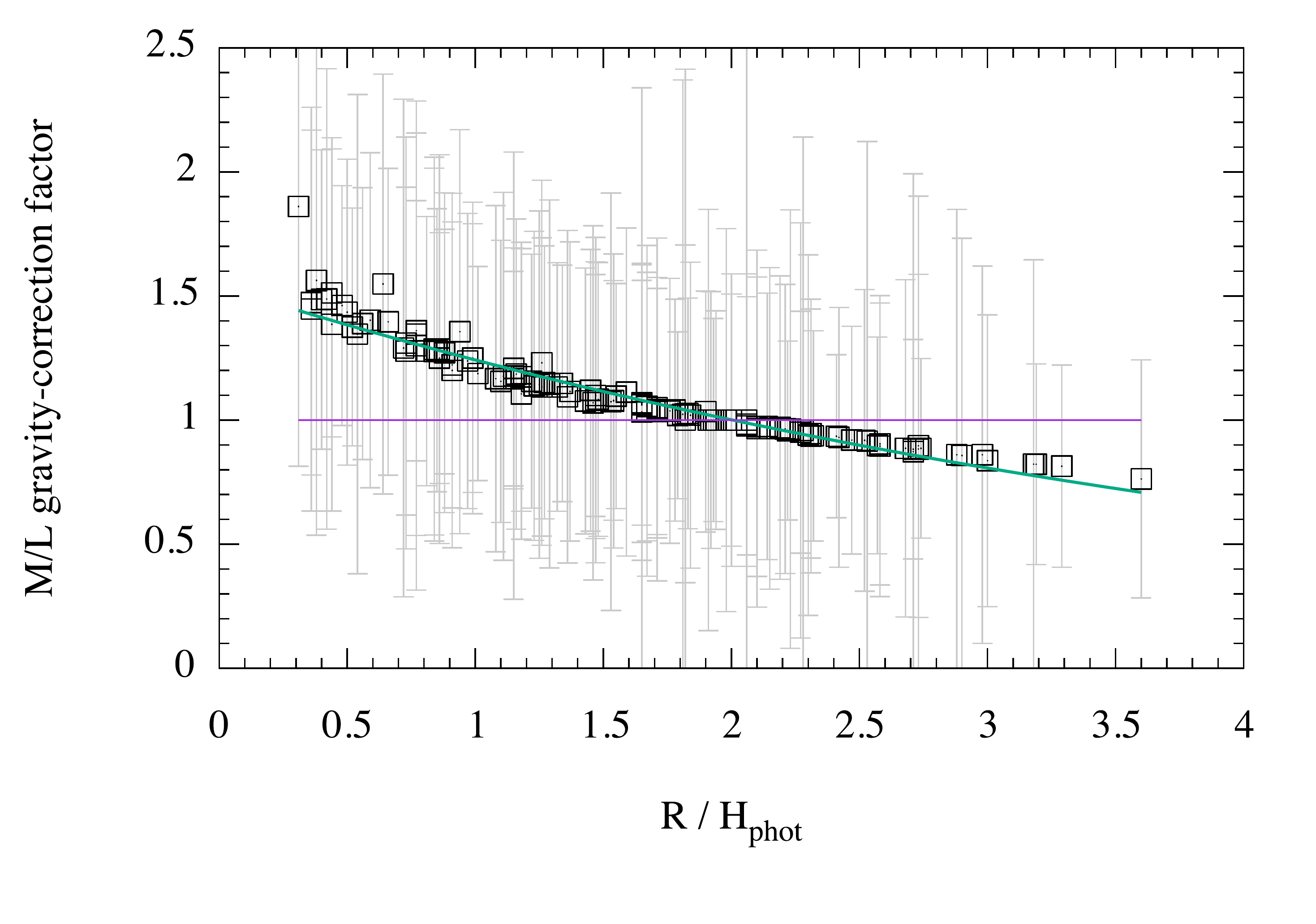}
\caption{The relative changes in the resolved DiskMass stellar M/L from Martinsson et al. (2013) solely due to the gravity correction (first term in Eqn.\,\ref{eqn:Sigma*}).  The error bars include Martinsson et al.'s estimate for systematic errors but do not include the much smaller effects of the errors in $H_*$ and $h_*$.  Also shown is an exponential fit (green line) used in the text.}
\label{fig:m2l}
\end{figure}

One can estimate the effects of the DiskMass assumptions on the derived stellar M/L and the dangers of assuming $H_p \! \equiv \! H_*$ using the tabulated values of $h_*$ and $H_p$ in Table\,1 of 
\cite{2013A&A...557A.131M} 
-- the former derived using Eqn.\,1 in 
\cite{2010ApJ...716..234B} 
-- and the $\Upsilon_*(R)$ data read off the published graphs in their Appendix A.3.
Fig.\,\ref{fig:m2l} shows the relative changes in the radially resolved stellar M/L derived by 
\cite{2013A&A...557A.131M} 
{\it solely} when the gravity-correction using the formula in Table\,1 is applied, i.e. without any change in the effects of the ISM; these are the minimum necessary corrections to the published DiskMass M/L.
The error bars include their estimates for systematic effects, and the inner bulge regions were ignored.
Most of the points lie on a well-defined curve that declines with radius due to their assumed correlation between $H_R$ and $h_z$ ($\xi_*$ depends both upon $R/H_*$ and $h_*/H_*$).
Since Martinsson et al.~conclude that most of their $\Upsilon_*$ are consistent with being constant, the effects of this correction can be described by assuming
\begin{equation}
\Upsilon_*(R) \approx \overline{\Upsilon}_{*,DMS} \, \exp^{-(R-2.3 H_*)/H_\Upsilon}
\label{eqn:upsilon}
\end{equation}
where $H_\Upsilon$ is the radial scale-length of the M/L variations, and $R\! = \! 2.3 H_*$ is where $\xi_* \! = \! 1$ and hence $\Upsilon_*$ matches.
$H_*^{-1} \approx H_\Upsilon^{-1} \! + \! H_p^{-1}$ is then smaller than $H_p$ because $\Upsilon_*$ decreases with radius.
The typical ratios $H_\Upsilon/H_p$ and hence $H_*/H_p$ for the DiskMass data
\begin{equation}
\frac{H_\Upsilon}{H_p} \approx 4.63\pm 0.10,
~~~
\frac{H_*}{H_p} \approx 0.82\pm 0.01,
\label{eqn:scalelengths}
\end{equation}
(internal errors only) were used to calculate the displayed M/L correction factors in Fig.\,\ref{fig:m2l}.
The final effect is substantial: the stellar surface densities are increased by up to 50\% for $R < 2 H_p$.
This result should not come as a surprise: variations in $\Upsilon_K$ of this magnitude are easily explained by differences in metallicity and age; indeed, \cite{2001ApJ...550..212B} say
%
``{\it ... because color gradients are common in spiral galaxies, significant gradients in stellar M/L ratio should be present in most spirals, in the sense that the outer regions of galaxies will tend to have lower stellar M/L ratios than the inner regions of galaxies (assuming a universal IMF).}''

Having replaced Eqn.\,\ref{eqn:DiskMass} with Eqn.\,\ref{eqn:Sigma_dyn} and predicted radial variations in M/L, it would seem surprising that the radial scale-length of $\Sigma_{dyn}$ would still be close to $H_p$, the prediction of the simple Van den Kruit model.
From Eqn.\,\ref{eqn:eqn} and Figs.\,\ref{fig:xi} \& \ref{fig:m2l}, we can write
\begin{multline}
\Sigma_{dyn,*}(0) e^{-R/H_{dyn,*}}
	\approx 
	\left[ \xi_*(0) e^{R/H_\xi} \right] ~ \left[ \Sigma_*(0) e^{-R/H_*} \right] \\
	~~~ \approx 
	\left[ \xi_*(0) e^{R/H_\xi} \right] ~ \left[ I(0) e^{-R/H_p} ~ \Upsilon_*(0) e^{-R/H_\Upsilon} \right]
	\label{eqn:scales}
\end{multline}
A crude fit to $\xi_*$ for a value of $h_*/H_* \! \approx \! 0.11/0.82 \! = \! 0.13$ typical of the DiskMass galaxies in the range $0.5 \! < \! R/H_* \! < \! 3$ yields $H_\xi/H_* \! \approx \! 7.3$ and so $H_\xi/H_p \! \approx \! 7.3*0.82 \! = \! 6.0$.
The scale-length for gravity variations is thus similarly large as those of the M/L variations, albeit with gradients of opposite signs.
The prediction for the stellar dynamical scale-length is then
\begin{equation}
\label{eqn:Hratios}
\frac{H_{dyn,*}}{H_p} \approx \left(-\frac{H_p}{H_\xi} \! + \! \frac{H_p}{H_\Upsilon} \! + \! \frac{H_p}{H_p} \right)^{-1}
	\approx 1.05
\end{equation}
%
%
which is easily within the error limits of $H_{dyn}/H_p \! = \! 1.17 \! \pm \! 0.25$ found by 
\cite{2013A&A...557A.130M}. 
The inclusion of the effects of $4/3 \Sigma_g$ with $H_g \approx 2 H_p$ would flatten $\Sigma_{dyn}$ relative to $\Sigma_{dyn,*}$ and the inclusion of the DM halo effects would result in considerable variation in general, again in agreement with the DiskMass result.

These effects also have relevance to the results of 
\cite{2014ApJ...797L..28S}, 
who found small values of $\Upsilon_*$ when extrapolating the kinematics to the disc centres.
By neglecting the $\xi_*$-effects, they assumed
\begin{equation}
\Sigma_{dyn}(0) e^{-R/H_{dyn}} \approx I(0) e^{-R/H_p} ~ \Upsilon_{dyn}(0) e^{-R/H_\Upsilon}
\label{eqn:swaters}
\end{equation}
(for an assumed constant vertical scale-height), i.e.
\begin{equation}
\frac{H_{\Upsilon,DMS}}{H_p} 
	\approx
	\left( \frac{H_p}{H_{dyn}} - 1 \right)^{-1} 
	\approx
	\left( \frac{1}{1.17} - 1 \right)^{-1}
	\approx
	-6.9
\end{equation}
Thus, their predicted M/L {\it increases} slowly rather than {\it decreases} with radius (a minus sign is missing in their Eqn.\,3), which is why their value for $\Upsilon_{dyn}(0)$ corrected for the radial dependence of the M/L is even lower than the uncorrected one.
They assumed that an average M/L ratio can be measured at $R \! = \! H_p$, but Fig.\,2 
shows that a more reliable value could be found at $R \! \approx 2.3\,H_* \approx 1.9\,H_p$ where $\xi_* \approx 1$.
Assuming that the DiskMass measurement of the dynamical surface density is roughly correct at this larger radius (i.e. even though using too little ISM and no correction for DM) and using Eqn.\,\ref{eqn:scalelengths}, one can easily see that the true value at $H_p$ is a factor of $\sim e^{-(1-1.9)/4.63}$ or 20\% larger:


\section{The effects of a thick disc}

\cite{2010ApJ...716..234B} 
discussed the possibility of having to deal with thin- and thick-disc components in some detail (their Section 2.2.4).
However, their discussion was mostly limited to the question of detecting the component in the surface photometry.
Since $\sigma_z^2 \propto h_*$ (Eqn.\,\ref{eqn:z13}), the larger vertical scale-height of a thick-disc automatically makes up somewhat for the lack
of surface density unless the relative contribution is so small that the effect is lost in the wings of the observed line-of-sight dispersion profiles.
Given DiskMass' typical instrumental spectral resolution of $10-20\,\KMS$ 
\citep{2010ApJ...716..234B} 
and a broad range of thin- to thick-disc dispersions in the local Milky Way between $10$ and $50\,\KMS$ 
\citep{2012ApJ...753..148B}, 
the danger of not being able to disentangle multiple components kinematically is real enough.
While 
\cite{2016MNRAS.456.1484A} 
discuss the kinematic effects of including a thick disc, they do so within the context of a purely stellar infinite disc.

Using our revised \cite{2013ApJ...772..108Z}, 
model, one can calculate the kinematic effects of separating the stellar disc into two thin- and thick-disc components with vertical scale-heights $h_t$ and $h_T$, respectively.
If $\Sigma_* = \Sigma_t + \Sigma_T$ is the total stellar surface density, one can define $\eta_T \equiv \Sigma_T/\Sigma_*$ as the fractional surface density of the thick disc.
In order to define observed quantities, one must introduce quantities weighted not by projected surface density but by the surface brightness $I$ via $\Upsilon \propto \Sigma / I$.
For instance, the weighted vertical scale-height measured at some fiducial radius $R_0$ -- presumedly equal to that used by DiskMass (Eqn.\,\ref{eqn:h*}) -- is then
\begin{eqnarray}
h_* & \! \equiv \! & \!
	h_*(R_0)  \approx
	\frac{I_t(R_0) h_t + I_T(R_0) h_T}{I_t(R_0) + I_T(R_0)} 
	\\ \nonumber
	& \! \approx \! & \!
	\frac{\Sigma_t \Upsilon_t^{-1} h_t + \Sigma_T \Upsilon_T^{-1} h_T}{\Sigma_t \Upsilon_t^{-1} + \Sigma_T \Upsilon_T^{-1}}
	\approx 
	\frac{1-\eta_T(R_0)}{u_t(R_0)} h_t + \frac{\eta_T(R_0)}{u_T(R_0)} h_T
\end{eqnarray}
where $\Upsilon_*$ is the $\Sigma$-weighted harmonic average of the thin- and thick-disc M/L
\begin{equation}
\Upsilon_*^{-1} \equiv \frac{\Sigma_t \Upsilon_t^{-1} + \Sigma_T \Upsilon_T^{-1}}{\Sigma_t + \Sigma_T} = (1-\eta_T) \Upsilon_t^{-1} + \eta_T \Upsilon_T^{-1}
\label{eqn:M2L}
\end{equation}
and where $u_t \equiv \Upsilon_t / \Upsilon_*$ and $u_T \equiv \Upsilon_T / \Upsilon_*$ are the relative thin- and thick-disc M/L, quantities which are presumedly not quite as dependent upon the details of stellar populations and star-formation histories as the absolute ones.

Finally, one can define the surface-brightness-weighted dispersion as
\begin{equation}
	\overline{\sigma_z^2}(R) \approx \frac{1-\eta_T(R)}{u_t(R)} ~ \overline{\sigma_{z,t}^2}(R) ~ + ~ \frac{\eta_T(R)}{u_T(R)} ~ \overline{\sigma_{z,T}^2}(R)
\end{equation}
so the composite dynamic surface density is
\begin{eqnarray}
\Sigma_{dyn} \!\!\!
	& \approx & \!\!\!
	\xi_t \Sigma_t \epsilon_t
	+ \xi_T \Sigma_T \epsilon_T
	+ \frac{16}{3} \rho_{DM} h_* \epsilon_{DM} 	
	\\ \nonumber
	& &
	~~~~ 	
	+ \frac{4}{3} 1.4 \left( \zeta_{HI} {\SHI} + \zeta_{H_2} {\SmH} \right) \epsilon_{ISM}
\end{eqnarray}
where
\begin{eqnarray}
\epsilon_{ISM}(R) \!\!\!
	& \! \equiv \! & \!\!\!
	\frac{h_t}{h_*} \frac{1-\eta_T(R)}{u_t(R)} + \frac{h_T}{h_*} \frac{\eta_T(R)}{u_T(R)}
	~~ = ~~ \epsilon_{DM}(R)
	\\ \nonumber
\epsilon_t(R) \!\!\!
	& \! \equiv \! & \!\!\!
	\frac{h_t}{h_*} \frac{1-\eta_T(R)}{u_t(R)} + \frac{4}{3} \frac{h_T}{h_*} \frac{\eta_T(R)}{u_T(R)} \left(1 - \frac{h_T^2}{(h_t+h_T)^2}\right)
	\\ \nonumber
\epsilon_T(R) \!\!\!
	& \! \equiv \! & \!\!\!
	 \frac{4}{3} \frac{h_t}{h_*} \frac{1-\eta_T(R)}{u_t(R)} \left(1 - \frac{h_t^2}{(h_t+h_T)^2}\right) + \frac{h_T}{h_*} \frac{\eta_T(R)}{u_T(R)}
\end{eqnarray}
are additional correction factors due to the presence of two stellar mass components.
Note that all quantities other than the scale-heights $h$, which we have assumed to be constant, are $R$-dependent.
so that the connection between the dynamic and stellar surface densities is now much more complicated: $\xi_t$, $\xi_T$, $\eta_T$, $\Upsilon_t$, and $\Upsilon_T$ are all different functions of cylindrical radius.

We can estimate typical values of $\eta_T$, $u_t$ and $u_T$ for the local Milky Way.
\cite{2008ApJ...673..864J} 
found $\rho_T(0)/\rho_t(0) \approx 0.12$, $h_t \! \approx \! 300\,\PC$, $h_T \! \approx \! 900\,\PC$ and hence $\eta_T \approx 0.26$ using dwarf M stars.
\cite{2016MNRAS.456.1484A} 
analysed the vertical velocity dispersions of local giants within roughly volume-limited regions out to $0.6-1.5\,\KPC$ above or below the plane of the Milky Way; assuming their mean-square-dispersion is a surface-density-weighted value, one obtains $\eta_T \approx 0.3$, in agreement with the value by Juri\'c et al.
Using a Besan\c{c}on model and an assumed star-formation history, Aniyan et al. estimate that $\Sigma_t \Upsilon_t^{-1} / \Sigma_T \Upsilon_T^{-1} \approx 1.56$, suggesting (Eqn.\,\ref{eqn:M2L})
%
$u_t \! = \! (1.56\!+\!1)(1\!-\!\eta_T) \! \approx \! 1.9$ and
$u_T \! = \! (1/1.56\!+\!1) \eta_T \! \approx \! 0.4$,
%
i.e. $(1\!-\!\eta_T)/u_t \approx 0.4$ and $\eta_T/u_T \approx 0.6$.
The surface-brightness-weighted scale-height is then $665\,\PC$ whereas the mass surface-density-weighted value is only $456\,\PC$
The factor of nearly 5 difference in $u_t$ and $u_T$ and the large different between the weighted scale-heights show that the relative M/L effects are indeed significant, at least for the local Milky Way.

Equally important are the effects of different radial scale-lengths:
the Milky Way's thick disc probably has a much smaller radial scale-length -- $H_T \! = \! 1.8\,\KPC$ versus $H_t \! = \! 3.4\,\KPC$ for the thin disc \citep{2011ApJ...735L..46B} --  which increases $\xi_T$ relative to $\xi_t$ as well.
If $R_0/H_t \! = \! 8.2/3.4 \! = \! 2.4$ and $h_t/H_t \! = \! 0.09$, then Fig.\,\ref{fig:xi} shows that $\xi_t \! \approx \! 1$.
The corresponding value for the thick disc with $h_T/H_T \! = \! 0.5$ and $R_0/H_T \! = \! 8.2/1.8 \! = \! 4.6$ (off the scale in Fig.\,\ref{fig:xi}) is $\xi_T \! \approx \! 2$.
If one assumes $H_* \approx (1-\eta_T) H_t/u_t + \eta_T H_T/u_T \approx 2.4\,\KPC$, the value for the mean disc is $\xi_* \! \approx  \! 1.2$.
An alien astronomer looking down on the Milky Way at the radius of the Sun would see an effective dynamic stellar and gaseous surface densities which are factors of 1.4 and 1.3-1.7 larger than the actual ones, respectively.
Thus, the effects of not including a thick-disc component can be severe, even when the contribution to the total surface density is modest.

Unfortunately, it would be difficult to detect the thin- and thick-disc components spectroscopically with moderate resolution spectral data like that used by DiskMass.
If the radial scale-lengths of the thin- and thick-disc components are different, one might be able to see this effect as a modification in the surface density profile; a significant thick-disc component with a significantly smaller radial scale-length should produce a Type III surface density profile 
\citep{2005ApJ...626L..81E}. 
However, 
\cite{2010ApJ...716..234B} 
checked for an effect and could see no evidence in the DiskMass data.

Barring a means of detecting the two components in individual galaxies, it may be possible to obtain statistical estimates for the relative contributions by comparing the results of many edge-on galaxies with those of face-on galaxies.
However, until a correction for a thick disc component can be made, this uncertainty will be a major source of systematic error in the stellar densities and hence derived M/L and DM halo parameters.


\section{Are the discs maximal after all?}

Galaxy discs are considered to be ``maximal'' if their contribution to $\V_{c}$ at $R=2.2 H$ is 85\% or more of the total 
\citep{1997ApJ...483..103S}, 
i.e. if the disc stars contribute $(0.85)^2$ = 72\% of the centripetal budget where their contribution should be maximal.
Although the DiskMass algorithms estimated $\Sigma_*(R)$ directly,
the final DiskMass models were first used to calculate $\overline{\Upsilon}_*$ and then to use the IR surface brightness to get the final values for $\Sigma_*(R)$ used to determine the stellar contribution to the rotation curve.
\cite{2013A&A...557A.130M} 
found values less than $(0.75)^2\!=56$\% for the disc baryons at $R=2.2 H_p$:
to achieve maximality, one would need to increase the effective baryonic surface density by about $0.72/0.56$ or 34\%.

In the previous sections, I have shown how the re-interpretation of the dynamic surface densities, the inclusion of a more realistic vertical gravity, the inclusion of more gas, the correction from $H_p$ to $H_*$, and the presence of a thick disc changes the derived baryonic surface densities relative to those estimated via the DiskMass assumptions.
Without access to the full DiskMass dataset, it is difficult to estimate the impact of these corrections on, e.g., the implied DM halo models -- each galaxy has its own properties and the corrections can result in more or less baryonic mass.
However, for the Milky Way analogue galaxies dominated by the stellar component, we can forget the modified ISM contribution and use the gravity correction to the stellar surface density alone, in an attempt to determine whether the baryonic discs might be maximal after all.
By changing $\Upsilon_*$ from a constant to a roughly exponentially falling function of radius with scale-length $H_\Upsilon \! \sim \! 4.6\,H_p$, intersecting the DiskMass value at $\sim \! 2.3\,H_* \! \approx \! 1.9\,H_p$ (Eqs.\,\ref{eqn:upsilon} \& \ref{eqn:scalelengths}), the total stellar mass $\M_*$ of the disc 
\begin{equation}
2\pi \! \int_0^\infty \! \Sigma_*(0) e^{-R/H_*} R dR 
= 2\pi \Sigma_*(R_0) e^{R_0/H_*} H_*^2
\end{equation}
is increased by a factor of
\begin{equation}
\frac{\M_*}{\M_{*,DMS}} 
	\approx \left(\frac{\Upsilon_*(1.9 H_p)}{\overline{\Upsilon}_{*,DMS}} \right)
			\left(\frac{e^{1.9/0.82}}{e^{1.9}}\right)
			\left(\frac{H_*}{H_p}\right)^2 
	\approx 1.01
\end{equation}
i.e. is essentially unchanged: the first term is $\sim\! 1$ by construction, and the the heightened inner surface density is compensated by the smaller scale-length.
Similarly, the relative change in the rotation velocity contribution of the stellar disc
\begin{equation}
\V_c (R)^2 \approx \frac{G \M_* R^2}{2 H_*^3} \chi\left(\frac{R}{2H_*}\right),
~~
\chi(x) \equiv I_0(x) K_0(x) \! - \! I_1(x) K_1(x)
\end{equation}
\citep{1970ApJ...160..811F} at $R = 2.2 \! H_p \! \approx 2.68\,H_*$ used by DiskMass to measure the ``maximality'' of the disc is easily estimated:
\begin{equation}
\frac{\V_c}{\V_{c,DMS}} 
	\approx
	\sqrt{ \frac{\M_*}{\M_{*,DMS}} \frac{2.68^2}{2.2^2} \frac{H_p}{H_*} \frac{\chi(2.68/2)}{\chi(2.2/2)} }
	\approx 1.09
\end{equation}
i.e. is only 10\% higher than for the DiskMass case.
The result is also only slightly higher if one measures the disc contribution at $2.2\,H_*$ instead of $2.2\,H_p$.
Thus, while the corrections to the DiskMass algorithms definitely result in major changes in detailed galaxy properties like mass-densities and scale-lengths, these changes are such that they roughly cancel out kinematically, producing the same centripetal contributions and resulting in the same ``maximality''.
Since the corrections for the ISM will only reduce the stellar contribution and are unlikely to result in larger baryonic fractions for $R < 2.2\,H_*$ -- most of their gaseous mass is in their outer regions 
\citep[see their Fig.\,3]{2013A&A...557A.131M} 
-- these results are unlikely to be changed even for gas-rich normal galaxies.
As we will see in the final section, however, the inclusion of the DM halo must result in different global masses, an effect which could increase the baryonic disc mass after all.

\begin{table}
\caption{Median DiskMass Galaxy \& Model Parameters}
\begin{tabular}{lll}
\hline
\multicolumn{2}{l}{Parameter} &	Value \\
\hline
$L$			&	\! radial scale-length of $\sigma_z^2$	&	$3.9\,\KPC$ \\
$M_*$		&	\! total stellar mass					&	$2.1\,10^{10}\,\MSUN$ \\
$H_p\!\!=\!\!H_*$	&	\! stellar radial scale-length			&	$3.9\,\KPC$ \\
$h_*$		&	\! stellar vertical scale-length		&	$0.46\,\KPC$ \\
$M_g$		&	\! total ISM mass						&	$1.0\,10^{10}\,\MSUN$ \\
\hline
$H_g$		&	\! radial scale-length of ISM			&	$7.8\,\KPC$ \\
$h_g$		&	\! vertical scale-height of ISM			&	$0.1\,\KPC$ \\
$\zeta_{ISM}$	&	\! ISM surface density correction	&	$1.3$ \\
$\rho_0$	&	\! central density of compact halo$^a$	&	$3.3\,10^8\,\MSUNCKPC$ \\
$R_c$		&	\! core radius of compact halo			&	$1.3\,\KPC$ \\
$\rho_0$	&	\! central density of extended halo$^b$	&	$6.0\,10^6\,\MSUNCKPC$ \\
$R_c$		&	\! core radius of extended halo		&	$10\,\KPC$ \\
\hline
\end{tabular}
{\small Case I$^a$ \& II$^b$ of Martinsson et al. (2013; Table\,5)} \\
\end{table}

\section{Discussion}

The task of modelling the stellar kinematical properties of spiral discs is a difficult one.
The DiskMass consortium went through an enormous effort to minimise the experimental errors and to propagate them correctly through the standard kinematic model, but the above sections have shown how important it is to define a proper kinematical model and difficult it is to constrain the kinematical parameters.
The kinematical effects don't always go in the same direction -- e.g. adding more gas increases the baryonic mass but also reduces the stellar mass needed to maintain a given vertical dispersion -- so it is difficult to say how these effects will play out in any given galaxy and what the radial distribution of these effects will be.

\begin{figure}
\includegraphics*[width=\columnwidth, trim=0 6.23cm 0 0.5cm]{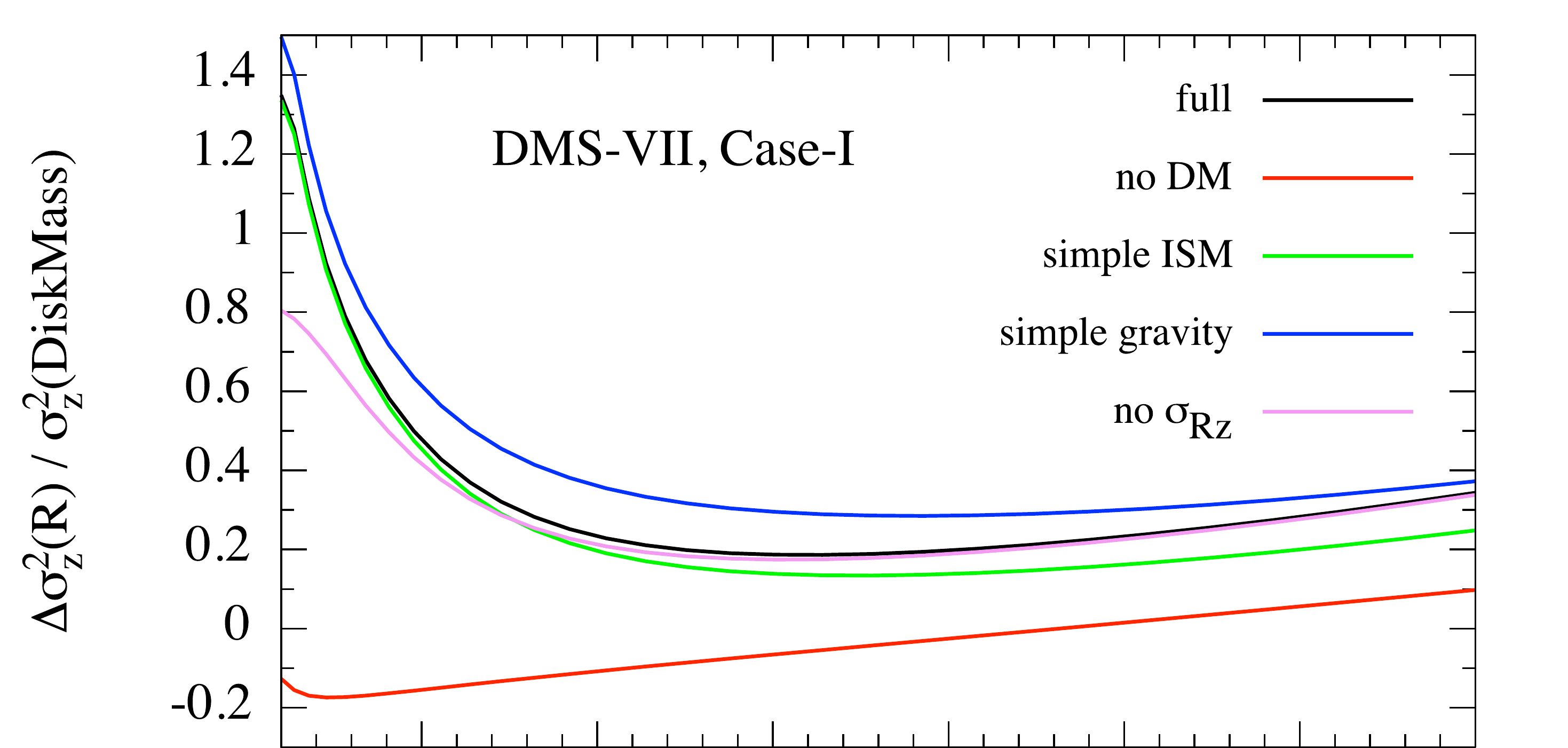} \\
\includegraphics*[width=\columnwidth, trim=0 2.0cm 0 8.35cm]{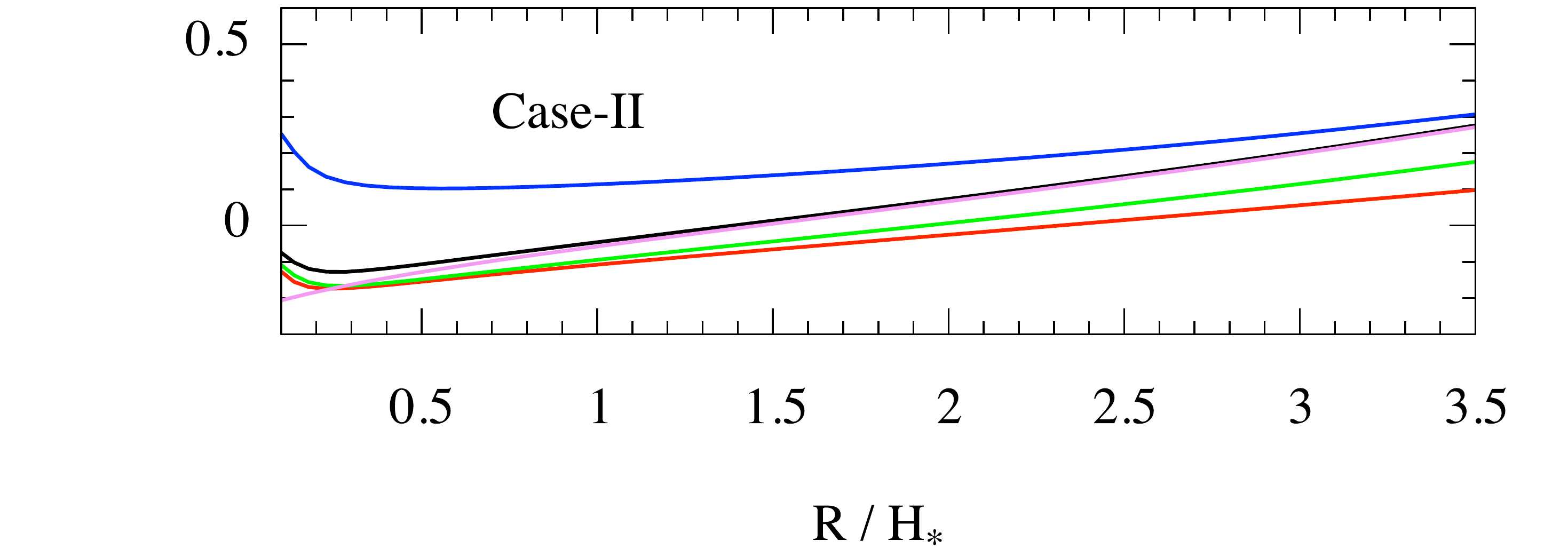}	
\caption{The relative change in $\overline{\sigma_z^2}(R)$
compared with that produced by the DiskMass model for a typical DiskMass galaxy (Table\,2) using the ``full'' model presented here (black lines), the full model minus the effects of DM (red), the full model but with the DiskMass ISM contributions (green), neglecting the effects of realistic gravity (blue), and the full model minus the $\sigma_{Rz}$-effects (violet).
The compact ``Case-I'' DM halo by Martinsson et al. (2013b) is used in the top panel, the extended ``Case-II'' DM halo in the bottom.
\CHANGES{The relative difference is a measure of the internal self-inconsistency of the simpler model.}
}
\label{fig:models}
\end{figure}

In an attempt to show how the DiskMass results might be affected for a ``typical'' DiskMass galaxy, one can consider a toy galaxy with the median baryonic properties shown in Table\,2 derived from Table\,3 in 
\cite{2013A&A...557A.131M} 
and a median isothermal DM halo corresponding to Cases I (compact halo) and II (more extended halo) from their Table\,5.
The resulting relative changes in the values of the modelled $\overline{\sigma_z^2}(R)$ are shown in Fig.\,\ref{fig:models} for a variety of cases and for both of their DM halo models: 
the ``full'' model displayed in Fig.\,\ref{fig:models} uses the complete analytic solution for $\sigma_z^2(z)$ by 
\cite{2015A&A...579A.123H} 
(the $\sigma_{Rz}$ cross-term model parameters $\eta \! \equiv \! 0$ and $\gamma \! \equiv \! \beta^{-1}\!-\!1$) integrated numerically over all heights $z$ and weighted by the assumed exponential vertical density distribution;
the ``no DM'' solutions ignore the DM term in Eqn.\,\ref{eqn:Sigma*};
the ``simple ISM'' solutions uses the DiskMass model for the kinematic effects of the ISM (Eqn.\,\ref{eqn:DiskMass});
the ``simple gravity'' solutions ignores the effects of realistic gravity ($\xi\!=\!1$); and the ``no $\sigma_{Rz}$'' solutions are equivalent to using $\beta\!=\!1$ (Eqn.\,\ref{eqn:sigmaRz}).
\CHANGES{A positive(negative) difference means that the improved model implies more(less) dispersion than would be found using the simplistic DiskMass kinematic model when using the same galaxy properties.
This is not a ``correction'' in the sense that the stellar masses used by the improved models are not those internally consistent with the actual dispersions.
Since the simple kinematic model assumes that $\overline{\sigma_z^2}(R)$ is directly proportional to $\Sigma_*(R)+\Sigma_{ISM}(R)$ only, and is effectively independent of the
DM halo properties (whose density is derived from the rotation curve only), $\Delta \overline{\sigma_z^2}/\overline{\sigma_z^2}(DiskMass)$ is rather a measure of how inconsistent the simplistic model is.}

The main effects are obviously the addition of a DM halo and the use of a realistic gravity, \CHANGES{though the corrections have opposite signs and the two DM halo cases are very different.}
The cuspy ``Case-I'' halo produces a very significant effect in the inner disc not included in the DiskMass models: unless one is willing to decrease the inner stellar surface density by a factor of $\sim\!2$, it is not possible to invoke a self-consistent cuspy halo.
\CHANGES{This is an important new constraint, since the halo properties are normally derived using the rotation curves only.}
The much smaller central density of the cored ``Case-II'' halo results in a decrease in the modelled $\overline{\sigma_z^2}$ whether or not the halo is included or not, requiring $\sim 10$\% more stellar mass. 
The radial form of the Case-II corrections are dominated by the gravity effect: 
this gradient could be reduced by decreasing the scale-height of the stellar disc (Fig.\,\ref{fig:xi}), which would also result in an increase in the baryonic M/L 
\citep{2016MNRAS.456.1484A}. 
Thus, both halo models imply that significant corrections are needed, albeit in very different directions.

The effects of the $\sigma_{Rz}$ cross-term are negligible for $R \! > \! H_*/2$, as expected, but increase dramatically at very small radii.
Thus, \cite{2016A&A...585A..17A}'s 
conclusion that there is an effect at this level is correct, but their model produced a {\it decrease} rather than an {\it increase} in the dispersion and the effect is only visible in integrated dispersions within the very inner disc.

Note that the corrections shown in Fig.\,\ref{fig:models} are for a given set of galaxy parameters, whereas the normal procedure is to fit the observations to determine those parameters.
Thus, it is clear that a re-analysis of the DiskMass data would result in very different derived galaxy properties, especially for the DM halos.


\section{Summary}

The use of stellar kinematic data in the analysis of spiral galaxy mass models represents a significant advance towards the goal of obtaining robust baryonic and DM mass profiles.
However, the dramatic increase in the amount and quality of available kinematic information requires that the traditionally simple dynamical model be replaced by a much more complicated one.
\CHANGES{Here, I have presented relatively simple but now self-consistent extensions to the standard model, including all of the effects of a realistic gravity field, higher ISM densities, the presence of a DM halo, and especially a thick stellar disc.
The extended model predicts that the stellar M/L ratios are generally not flat but decline gently as expected from the observed colours}, resulting in a radial surface density scale-length that is smaller than the photometric one by $\sim 80$\%.
\CHANGES{The inclusion of the effects of the DM halos on the stellar dispersions makes it much more difficult to maintain cuspy DM halos traditionally only constrained by the rotation curves.}
Exactly how a more complicated model will change the fitted stellar M/L, the ``maximal-ness' of the baryonic discs, and the DM halo properties for any given galaxy remains to be seen in detail, but it is clear that the differences will be significant.


\section*{Acknowledgements}

I would like to thank S.\,McGaugh for suggesting this paper and especially the referee, M.\,Bershady, for a critical reading of the original manuscript that resulted in considerable improvements.



\bibliographystyle{mnras}
\bibliography{diskmass}

\bsp	
\label{lastpage}

\end{document}